# Intuition in Einsteinian Physics

Bernard Schutz

> **What?**
> This chapter sets the stage for the rest of the book by exploring the role of intuition as a tool to deepen understanding in Einsteinian physics. Drawing on examples from the history of general relativity, Bernard Schutz argues that the development of physical intuition is a crucial goal in physics education in parallel with any mathematical development of a physics subject.
>
> **For whom?**
> Readers who wish to learn how expert physicists think conceptually about their subjects to understand them. Readers who wish to see how we can introduce Einsteinian physics to students by developing their intuition as well as teaching them the mathematics.

## Introduction

If you say the word "Einstein" to an average person, the first thing likely to come to their mind is an image of his iconic wild hair. The next thought will probably be that his science, however important, is far beyond the ability of the average person to understand. Both of these things place Einstein outside everyday society, a curious phenomenon, to be wondered at. The teaching of physics at schools has until recently reinforced this view, mostly leaving Einsteinian physics to specialist university courses. The mathematical framework for this physics is challenging, and that has seemed a good reason to teach it only to those advanced university students who have mastered enough mathematics.

But it can't be left to the boffins any more. Everyone has heard of black holes; even small children want to know what they are. Recently, gravitational waves have also entered our vocabulary: their first detection in 2015 caused a media sensation, coming as they did from *two* black holes *merging together!* Other terms in common use that could come straight out of a general relativity textbook include the big bang, wormholes, and even the warp drive. And then there is the quantum side of Einsteinian physics. Our lives would come to a halt without the internet, computers, mobile phones, microwave ovens, and other devices that only work because of quantum principles. I would guess that well more than half of the world's economic activity today is enabled by physics that was only developed in the last 100 years and which is still thought to be too difficult to teach at the school level.

Does this leave us in a tough spot? Is the world to be divided between "users" who have no idea what is inside the box, and "experts" whose work is too arcane to explain to the rest of the world? That would be dangerous for society. If we don't want that, can we open up this world of physics concepts to people who don't have the mathematics to deal with the equations in which the physics is framed?

I believe that we can, and that we can at the same time improve the way we teach Einsteinian physics even at the advanced level. The way to do it starts with realizing that even the experts think conceptually about their subjects in order to understand them; they don't go around just solving complex equations any time they want to answer a question. They use what we call *physical intuition*: a basic understanding of how things work and fit together that requires little or no mathematics. We should be introducing Einsteinian physics to students by developing their intuition as well as teaching them the mathematics.

Without intuition, science just would not work. Scientists today are specialists, and communication between different specialties requires a shared physical intuition. For example, if an astronomer wants to understand how a black hole might affect a star that is near it because he might want to observe the star in his telescope, he does not need to be able to solve Einstein's equations himself. Rather, he can approach a specialist in general relativity for insight, and she will explain to him how the gas gets pulled off the star by the black hole's gravity, how it will then get hot and therefore bright as it gets near the hole, and finally how it will vanish from view after being swallowed by the hole. None of their conversations will use arcane mathematics. Instead, they will rely on shared intuitive concepts about fluids and heat and gravity. And: these are concepts that are taught in school!

What is not always taught, either at school or at university, is an appreciation of the important role played by intuition in science. Some physicists I know don't display much intuition; they work mostly with the mathematics of their specialty, and they consequently don't have much interaction with other specialties. Other physicists have been famous for their intuition. One example was Richard Feynman, who won the Nobel Prize for helping develop the theory of quantum electrodynamics. He was a member of the scientific panel that investigated the deadly 1986 Challenger disaster, when a space-shuttle launch failed because its booster rockets exploded. Feynman famously demonstrated on television that the fault lay with the decision to launch in cold weather; he dipped a piece of rubber used in the booster assembly into ice water, and then simply snapped it in half. End of discussion. There are plenty of other stories showing how Feynman was able to cut through confusions in apparently complex physics problems with a simple physical argument that even a high-school student could understand.

It is important to be clear about what intuition is. I would distinguish between the kind of intuitive thinking Feynman demonstrated and "popular science" presentations, although there is obvious overlap. Popularizing science is important, but such talks often aim at entertaining the audience. Here we aim at teaching young scientists to use intuition in order to *do* science more effectively.

I believe that intuition plays a key role in scientific thinking, even though in casual discourse, intuition is often used as a synonym for sloppy thinking. Intuition is not the opponent of logic. It is its complement. A brain governed entirely by logic would be a brain in paralysis, unable to generate an original thought. We educators need to guide students toward a fruitful interplay between the creativity and model-building of the intuitive process, and the analytical and critical facilities of logic.

I will, therefore, focus in this chapter on *developing* physical intuition as part of learning physics, and then using intuition as a tool for deepening understanding, one that works in parallel with any mathematical development of a physics subject. The teacher can then grab hold of those intuitive concepts in students' minds to link up with other parts of physics, even parts where the mathematics is more complex. This should be done at school (and it often is) and university (often not). Doing this is important because, as I will argue later, intuition is *central* to how our brains work, how we make sense of the real world. We are not computers, and teaching us physics is not like programming a computer. So my aim in this chapter is to explore how we can use the way intuition works, in physics and in the physicist's brain, to improve the learning of science.

# Intuition in General Relativity: Examples

First, I will try to make the notion of intuition more concrete by giving three examples from Einstein's most mathematically complex work: general relativity. The first example is from the history of the subject. The equations were more or less complete by late 1915 (the cosmological constant came later), the fundamental solution for what we now call a black hole was discovered by Schwarzschild in 1916, Einstein discussed gravitational waves already in 1916 and arrived at what we now call the "quadrupole formula" showing how systems emit gravitational waves in 1918. In the 1920s, the theory was applied to cosmology, ready for Hubble's 1929 discovery of the expansion of the universe. So one might have expected rapid progress following on from these major achievements, placing general relativity on a firm physical basis. But instead we find that, even into the 1950s, physicists could not agree on whether Schwarzschild's solution described real objects, nor on whether gravitational waves were physically real or just an artefact of the complicated mathematics. Einstein himself changed his mind on these questions several times.

In my view (Schutz 2012), the root problem was that, in the period 1930-1950, people who worked on relativity approached it mainly mathematically. Naturally, we don't come into the world equipped with a relativity intuition: we have to gain it through "experiences" in the subject. Scientists commonly do this when entering unfamiliar territory using an intuitive tool called heuristics, in which complex systems are approximated by a few simple characteristics, that might be found by studying particularly simple examples. The mathematics of relativity gets quite complex for realistic problems, and occasionally people (including Einstein) wrote papers with fundamental errors that simply clouded the subject. But in the early period, few of the mathematicians working in relativity tried to work on heuristics; instead, they trusted only what they could get mathematically from the full Einstein equations.

In the 1950s real progress in relativity began because some brilliant physicists who were already skilled in heuristic thinking began working in the field. A famous example was Feynman himself, who at a key conference in 1957 (DeWitt & Rickles 2017) gave a simple physical argument showing that gravitational waves had to be real and had to carry energy (up until then hotly questioned). Feynman also pointedly remarked at this meeting that progress in physics did not always come from doing the mathematics better, but could come instead by working on simple situations or approximations, and then by doing experiments. What he was saying was that relativists needed better physical intuition.

Physicists spent the next several decades developing this intuition: working simple examples, using approximations to bridge between the simple examples, asking physical questions about what could be measured by experiment and observation, and doing the experiments and observations. The 1957 conference had kicked off discussions aimed at intuition-building among a growing number of scientists working in this field. The more scientists joined in, the more important it was to develop a language, a body of heuristics, an intuition that could be shared, that would help them communicate with one another.

By around 1990 there was a well-developed corpus of heuristics about black holes, gravitational waves, gravitational collapse, and cosmology, the study of which led relativists to a deeper intuitive grasp of the theory. In parallel, astronomers had been opening up a fascinating universe, beginning to gather evidence for black holes, gravitational waves and the Big Bang. This input of data nicely validated the heuristic concepts that the theorists were developing. At this point, 75 years after its auspicious beginnings, general relativity was at last ready to become a working part of astrophysics. Mathematics alone had not been enough to get there.

My second example follows one part of the above story. It is how physicists finally came to terms with black holes. We now understand that the Schwarzschild solution represents a black hole, deep inside of which is a "singularity", a place where gravitational forces become infinitely large; but that surrounding the singularity is a surface we call the horizon, which is a one-way membrane: things from outside can cross to inside, but not the other way around. We are troubled by the singularity because the laws of physics as we currently understand them don't tell us what happens when an infalling body reaches it, nor what happens next. But this is something we can live with because it is not even in principle observable: no information can get out to us through the horizon. Crucially, therefore, it does not disturb our ability to use physics to describe things outside the black hole. Notice that the description I have just given is intuitive: without mathematics you can still make sense of it because I have linked it with your understanding of causality in the physical world.

But it took a very long time for physicists to get to this intuitive picture. One reason is that it seems, from Schwarzschild's equations, that on the horizon time stands still. This was not consistent with the notion of causality that early 20th century physicists had. But early work by Painlevé, Gullstrand, and Lemaître had shown by 1935 that if you describe the Schwarzschild solution in what might be called "co-moving" coordinates – coordinates that are dragged along with in-falling particles – then time does not stop at the horizon, space is smooth and regular there, gravity is not infinite, and the particles just keep on going inwards. So the horizon was not a singular place, and the apparent problem with time was caused by the coordinate system Schwarzschild had used. Moreover, only a few years later Oppenheimer & Snyder (1939) had argued (using part intuition, part heuristics, and part mathematics) that a sufficiently massive collapsing star goes all the way down to its "gravitational radius", which is where we now understand that it turns into a black hole, leaving the horizon at that location. Convincing as these results seem to us today, they did not convince other people working in the field at the time, including Einstein, to take black holes seriously.

The final step wasn't taken until almost 20 years later, in 1957. John Wheeler, a distinguished quantum field theorist (he had been Feynman's PhD supervisor), felt it was time for relativity to be turned into a part of physics. To him, the most formidable problem was whether the Schwarzschild solution described real objects. A key issue that nobody had yet addressed was that the solution was perfectly spherically symmetric, and real bodies simply aren't. This symmetry was an idealization that had enabled Schwarzschild to simplify the Einstein equations and get an exact solution in the first place. Schwarzschild's solution was, therefore, a heuristic, a simplified model of the real thing. So Wheeler asked, how faithful to the real world is this model? If a real star collapsed, would it really become like the Schwarzschild solution, the way Oppenheimer and Snyder had argued?

It is said that Wheeler hoped that the answer would be no, that any small initial asymmetries of the star would get larger and larger as it collapsed, so that by the time it reached its gravitational radius it would fragment or do something else that would prevent it getting any denser, and prevent the formation of the horizon. So he set to work with Tullio Regge to prove this in the simplest way: study mathematically what happens to very small non-spherical changes in Schwarzschild's gravitational field. Wheeler hoped to prove that these small perturbations would grow exponentially with time, the sign of an *instability*. If Schwarzschild's solution were unstable, nothing like it would exist in the real world.

But in fact, Regge and Wheeler (1957) found the opposite: all perturbations either died away exponentially or became constant in time. This meant that a non-spherical star could indeed collapse through its gravitational radius, and the horizon would afterwards simply settle down exponentially rapidly into something very similar to the Schwarzschild solution. The Schwarzschild model did indeed resemble objects that could exist in the real world!

Regge and Wheeler's work was still, of course, a simplified case, a heuristic itself that validated the original black hole heuristic. It was only an approximate solution of Einstein's equations, valid only for tiny perturbations. A full exact solution was beyond anyone's ability until we became able to do fully numerical simulations on supercomputers. But it was exactly the kind of step that Feynman had asserted (in the same year) was needed: do the simple calculations, and use them to fit the theory (in this case, of black holes) into the real world. And it convinced Wheeler: he became the field's biggest proponent of the reality of black holes in astronomy. Heuristics and intuition are powerful drivers of scientific investigation.

My third example regarding intuition is drawn from my own teaching. Many years ago, I wrote a textbook called *A First Course in General Relativity*, now in its third edition (Schutz 2020). It introduces the full mathematics needed for Einstein's equations, aiming to get students to the point where they can solve those equations and understand the physical meaning of the solutions. But a one-year university course has only a certain number of hours, and so if you do the mathematics, you are limited in the amount of heuristic/intuitive development you can include. I used the exercises in each chapter to help with this and stressed intuitive ideas in my lectures. But it raised a challenge: how far into general relativity can one go without much mathematics, relying much more on intuition and heuristics?

I worked for almost 20 years on this challenge, and by doing so greatly deepened my own intuition about the theory. The result was the book *Gravity from the ground up* (Schutz 2003), and it illustrates the point of this chapter, that one can genuinely *teach* students to understand a subject like gravity – starting with Galileo and going all the way to general relativity – using only pre-calculus mathematics. In fact, most of the mathematics in this book is reserved for "boxes" that interested students can attempt, or they can just accept the results, incorporate them into their thinking, and go on to what comes next. The pedagogical aim of writing this book was to find the right way to link new concepts with ideas that students already had learned at school. The fact that this was possible taught me how the complexity of the mathematics of general relativity tends to hide the continuity of physical concepts from Galileo, through Newton, to Einstein. Teachers of relativity would do well to bring this intuitive continuity to the front as often as possible.

In view of this, I threw away the usual (chronological) order of presentation of gravity theory, where one first discusses all of Newtonian gravity and only then goes on to Einstein. Instead, it made much more sense to me to treat Newtonian gravity from the start as a special case of Einsteinian gravity, and to immediately employ Newtonian concepts, as soon as they are developed, as a way to introduce their Einsteinian relatives.

This approach starts right away in the first two chapters (there are 27 in all). Galileo's experiment of dropping balls from the Leaning Tower of Pisa is described in the first chapter, and Newton's extension of Earth-bound gravity to the whole solar system is treated in the second. Now, Galileo's experiment is historically the first statement of what we call the equivalence principle, which in turn is a foundation stone of general relativity. So Chapter 1 already explicitly lays that part of the foundation. And in Chapter 2, we re-express the equivalence principle in Einstein's form: an observer who is falling freely in a gravitational field measures no local effects of gravity at all, she is weightless. This allows us to prove in this chapter that light is redshifted as it climbs up out of the Earth's gravitational field. Einstein regarded this gravitational redshift as the most important prediction of all from general relativity. So here we have Galileo, Newton, and Einstein arm-in-arm. Any intuition that students may already have about Newtonian gravity begins building their relativistic intuition.

To me, this approach makes a lot of sense. The "weird" concepts of general relativity, like black holes, are more approachable if introduced alongside relevant parts of Newtonian gravity. Black holes themselves make their first appearance as early as Chapter 4, where I describe the 18[th]-century discussion of stars that are so massive and compact that they can in principle trap light, something that follows from Newtonian gravity and the equivalence principle. Students can take away from this that, when they later learn about black holes, their trapping of light is not something new with Einstein; it is a concept they can comfortably carry over from Newton. The dark Newtonian star is a useful heuristic on the road to Einsteinian black holes.

Intuition and heuristics not only help students learn; they guide good physicists in opening up new ideas. The 18th-century discussion of hypothetical dark stars is an example. Another remarkable example is the investigation in the 19th century by the famous mathematician Laplace of the possibility that gravity had a finite speed of transmission (not allowed by Newton), which Laplace realized would lead planets to very gradually spiral in toward the Sun (Laplace 1839). We know from Einstein that this is indeed a consequence of the emission of gravitational waves, which themselves are a consequence of the fact that relativistic gravity has a finite speed (that of light). Although the in-spiral effect in the Solar System is unmeasurable, it is today being observed in every gravitational wave detection. In fact, these detections beautifully combine these two Newtonian, pre-Einsteinian concepts: the waves come from black holes spiralling together. So, while these early speculations about dark stars and in-spiralling planets did not directly lead anywhere, they show that Einstein's general relativity emerged organically from some of the physics we teach at school. So it stands to reason that students at school who are developing their physical intuition should also find much of Einstein's physics natural.

## What is intuition, and how does it relate to mathematical logic?

Physics is often presented primarily as a framework of laws, which are expressed in the language of mathematics. If laws make experimentally verified predictions in their respective domains of validity (and we don't yet have a "theory of everything"), then in this picture we "understand" that domain. But is this really what understanding a domain means? It is true that, if we can calculate the mathematical consequences of a theory, then we can predict experimental and observational outcomes. When LIGO made its first detection of gravitational waves (Abbott, et al 2016), the observed waveform fit very well prior predictions, obtained using elaborate analytical calculations and supercomputer simulations, of waves from black-hole in-spiral and merger. This gave us confidence that the source really was a black-hole binary system. It validated our understanding of the way two such bodies orbit and merge, for example. But does our *understanding* consist just of the outputs of mathematical and computer calculations?

Physicists work with what we call models. Models can be elementary, even oversimplified. There is a common joke about the theorist who starts out saying "Let's consider a spherical cow"; it captures the danger of too much simplification. But even our deepest theories are themselves models, since we know they don't describe everything out there. And even if they did, our brains could not take it all in.

Models match how our brains work. Animals use their brains to model the outside world. The models work because they include what the animal needs for survival but they exclude irrelevant stuff. And the animal's model of the outside world itself includes and fits together a number of sub-models, which for example may handle the information from different sensory systems. In much the same way, our model of a black hole binary system contains sub-models of the two individual black holes. These models exclude irrelevant information, such as how and when the black holes were formed. When the binary separation is large enough, we can compute the orbit from Einstein's equations without knowing anything about the holes except their positions, masses, and spins. Our model of a single black hole tells us that ignoring its detailed structure at this point is fine. When the holes get closer to one another, we have to replace this simple quasi-Newtonian model (technically called a post-Newtonian model) with a model based on a full numerical simulation that uses Einstein's equations to compute how holes interact with each other, deform, and merge. Both models tell us what the gravitational waves emitted by the systems will look like. Based on them, we have another model that computes how these waves travel through the universe to our detector. That model excludes irrelevant things, like what happened to the binary system itself. And then we need another model for our detectors' response to the incoming gravitational waves. This model ignores where the waves came from or how they got here. The detector physics model is itself, of course, a hugely complex assemblage of models of the different parts of the detector. The models for different parts may be using the laws of physics of different domains, like surface physics or the quantum theory of light.

I understand some of these models, but as a theorist of general relativity, I don't understand many of the detector model's sub-models. Yet I will assert that I understand detectors and how they work. This is because I have a greatly simplified model of the detector in my head, but one which is faithful enough for my purposes. And that is the key – my purpose in understanding the detector is to know that I can rely on it to tell me about the gravitational waves, which are what I am really interested in. My colleagues who are detector physicists have a rather different set of objectives, which require a much more sophisticated detector model than mine. But for each of us, the model we have in our heads captures much more closely the notion of *understanding* than does simply our ability to write down the laws of physics that support and validate these models.

I believe that these models form the basis of our physical intuition. When a group of detector physicists want to understand what is going wrong with a detector, they do not immediately go back to first principles and begin solving a lot of fundamental physics equations. They first use their sub-models of different parts of the detector to try to pin down which sub-model might capture the domain where the problem originates. Then that sub-model is taken apart, into its constituent sub-sub-models, and so on. The problem will be found in one such model or perhaps in the way that their mutual interactions have been (incompletely) modelled. At some point, some difficult calculations might be needed, or computer simulations, but rarely will these involve going right back to the fundamental laws of physics. Learning physics, therefore, is much more than learning the mathematical logic of the fundamental laws. It is learning to keep models in one's head and learning how to knit them together into a representation of the real world. An intuitive physicist can do this really well. Therefore a student who is learning physics needs not just the equations, but also the key heuristic models that can be put together to make the whole domain understandable. These key heuristics were lacking for the physicists who, in the period 1930-1950, were struggling to make sense of general relativity.

Not coincidentally, this kind of structuring of physical understanding reflects what happens in our brains when we learn and think. Neuroscientists, cognitive scientists, and philosophers have in recent years been learning a lot about how the brain "knows" things. Much still remains to be discovered, but what is understood today is germane to our discussion. Decision-making in our brains is done via processes that resemble the working of neural networks in our computers, where information (both sensory and memory) is filtered through many layers until a "best" response is triggered. The process resembles what scientists call "Bayesian statistics", where one's confidence in the value returned by some measurement is computed not just from the known uncertainties in the measuring procedure, but also from uncertainties one assigns to the assumptions one makes. These assumptions, called priors, could be things like environmental factors that might have affected the measuring apparatus, or they could include information from previous measurements of the same thing. In astronomy, for example, they might include one's confidence in what astronomers already believe they know about the system being observed.

The brain makes decisions in what I would call a quasi-Bayesian way, relying on memory and hard-wired "instincts" for relevant priors and their uncertainties, and relying on senses or other mental events to trigger the need for a decision and to input new information that is relevant. The different neural-network layers apply tests of what is relevant, what is important for the decision. But where Bayesian theory would lead to a range of possible measured values with their probabilities, the brain is constructed to arrive typically at a single decision, which hopefully is the best for the animal in the circumstances. The way brains make decisions, therefore, is not deductive. It is fundamentally probabilistic, and it is only as good as the information that is used for it, including the stored experience from relevant previous situations.

It follows that, when we are presented with a problem, ideas about how to address it percolate up from below, and this percolation is unconscious. At some point ideas become conscious, and there is considerable debate among neuroscientists today about what exactly consciousness means. But for our purposes here, it is enough to know that we rely on our "gut" level for input to our consciousness. In other words, much of what we "think" has been arrived at statistically, not deductively. This captures our intuitive way of understanding the world, and its statistical nature is not a reason to distrust it: decision-making has been well adapted by evolution to deal with both situations where a lot is known and ones where little information is available to guide decisions. This point has been strongly argued, for example, in the book *Gut Feelings: the Intelligence of the Unconscious* (Gigerenzer 2007).

Even more challenging to the primacy of logic in human thinking is the recent book *The Enigma of Reason* (Mercier & Sperber 2019). They argue that logic and reasoning are useless to animals that cannot communicate about "how" they arrived at decisions: such animals simply follow their gut feelings, arrived at in this quasi-Bayesian way with inputs from built-in instincts, learning, and the senses. If the animals are part of a social group, then hierarchy and other externals determine which animal's decisions are followed by the group. But when language arose among humans, it became necessary to be able to persuade other people of one's own point of view, in order to preserve the working of the social group. So giving reasons became a way to influence others. Of course, then one had also to listen to other individuals' reasons, to respond to other individuals' challenges. A way had to be found to choose among conflicting decisions; the language of logic originated here. We have acquired the notion of logic through our cultural evolution. It is not biologically wired into our brains.

Simple as this line of reasoning may be, it has profound implications. It suggests that people normally do not arrive at decisions by exercising their faculty of reason, although of course, they can force themselves to if they are dealing with mainly objective facts (something that doesn't often occur in everyday life!). But even if they have taken a decision at gut level, believing they "know" it is right, people will invent reasons post-hoc if challenged. If they are part of a social group that all are inclined to the same choice (as much of politics seems to be in the age of the internet), then their reasons do not have to be particularly logical, since they won't be seriously challenged. Aristotle may have felt that the human brain was unique in being logical, but in the Mercier-Sperber view the human brain is like that of any other animal, except it is good at crafting justifications when it has to!

This allows us to reconsider how animals' brains use models of the real world around them. The purpose of such a model is to condition the way the animal reacts to new situations, new challenges, and help it to survive. We should think of these models, not as map-like representations of the world, but as algorithms for dealing with it. Intuitions about physics, in this view, can be seen as prescriptions for fitting different parts of physics together in the mind, prescriptions that help a physicist decide what to expect from a new experiment, decide whether to be comfortable or uncomfortable with a suggested explanation of a new phenomenon. They are particularly important in the creative side of physics, where new ideas are needed. Our intuitions are algorithms that guide the creation of these ideas, ensuring that we don't waste time on totally random junk. All of this works mainly in the unconscious mind, feeding decisions upwards to our conscious selves.

# Conclusion: Teaching physics - the interplay of intuition and logical reasoning

How do all these recent ideas about how our brains work affect our view of intuition and mathematics in physics? I have – and this is, of course, necessarily a simplification – identified intuition with the concepts/models that win the unconscious Bayesian competition among alternatives in our brains when we seek solutions to problems in physics. I would correspondingly identify mathematics as the language, and logic as the method, that we use post-hoc to justify the correctness of our intuition. Intuition is how we "know" science, and logical testing of intuition against experiment/observation, mediated by the language of mathematics, is how we make sure that what we "know" is "right".

As a social group, scientists put up a much stronger set of challenges to intuitions than one typically experiences in normal social activity. Competition among scientists, the insistence on validation by experiment or observation or mathematical calculation, and the tradition of refereeing research papers all ensure that ideas are challenged, that the proposer should not be able to get away with justifications that don't conform to high standards of logic. In particular, the ultimate appeal to objective experimental validation enables the whole community to agree on the way ideas are challenged and encourages scientists themselves to be more self-critical of their gut instincts, of their intuition, than is common in normal life. Politics and business, for example, operate on quite different rules. That is, I think, one reason why science is by-and-large successful in its own terms, why it normally advances its frontiers and rarely has to retreat and abandon territory it had claimed to understand. But its success does not come from relying purely on logic. It cultivates an especially self-critical form of intuition. But the interplay between intuition and logic is as much a part of science as of any other human activity.

So when physics is taught to students, the development of physical intuition must be a key goal. It needs to be cultivated. New concepts should be introduced using both intuition and mathematics, at whatever level is appropriate for the student. Explicit attention should be paid to "meaning". Difficult subjects, like Einsteinian physics, can be introduced successfully using intuitive concepts, and readers will find creative and exciting suggestions in later chapters of this book about how this can be done. Many of these intuitions will link up with ones that students already have, leading to a grasp of much of the physics even without elaborate mathematics. At school-level, most of the physics in these advanced subjects will be discussed at the intuitive level. But even in a university course, teachers should explicitly try to cultivate intuition and to use it to make the mathematics more understandable.

We are talking about scientific intuition here, which means it has to conform to the scientific method, has to be ready to withstand strong challenges from other scientists. Therefore the instruction needs to pay attention to the soundness of the intuitions that students are developing. Intuitions act below the level of consciousness, so it can be difficult for a teacher to know whether students are picking up useful or faulty intuitions. Teachers, therefore, need to create the kinds of challenges to intuition that scientific progress requires.

To ensure that students' intuitions are sound, students need to be self-critical about them, and they need to be able to defend those intuitions to others. Without introducing this kind of "quality control", for example using group discussions or individual student presentations, teachers risk leaving students with faulty concepts. We are good at assessing students' mathematical facility with a physics subject: we just give them written tests. But if we are going to guide the development of intuition, we need to get students to do what humans for millennia have done: justify their intuitions by explaining them to others, by giving convincing reasons.

Developing students' physical intuition can serve another important purpose, particularly because most of our school students won't become scientists. We want students to find science interesting, but we also want them to respect it. We want them to come away understanding how it deals with reality, how it makes progress. In a world full of fake news and reality-deniers, students need to develop their own scientific intuitions and to experience scientific debate in order to understand how working scientists "know" about the world. The acquisition of scientific expertise must not be seen as an arcane ritual, opaque to and hence untrusted by the general public. Paying attention to intuition gives us the opportunity to teach physics in an interesting way and to show all students that anyone can learn to think like a scientist.

If we want to accomplish this, then we should recognize that words and pictures and experiments matter just as much as equations, that argument and discussion matter just as much as written exams.

# Acknowledgements

I would like to acknowledge stimulating discussions with many of the contributors to this book, and especially to the editors M. Kersting and D. Blair. I am also grateful for insightful comments from C. Gorman.